\documentclass[12pt]{iopart}

\usepackage{iopams}
\usepackage[english]{babel}
\usepackage{graphicx}
\usepackage{cite}
\usepackage{color}
\usepackage{bbm}
\usepackage[normalem]{ulem}
\usepackage{times}
\usepackage[colorlinks=true,citecolor=blue,linkcolor=blue,urlcolor=blue]{hyperref}

\newcommand{\ket}[1]{\vert #1 \rangle}
\newcommand{\bra}[1]{\langle #1 \vert}
\newcommand{\ketbra}[2]{\vert #1 \rangle \langle #2 \vert}

\newcommand{\abs}[1]{| #1 |}

\begin{document}

\title[]{Parity-preserving light-matter system mediates effective two-body interactions}

\author{Thi Ha Kyaw}
\address{Centre for Quantum Technologies, National University of Singapore,
3 Science Drive 2, Singapore 117543, Singapore}
\ead{thihakyaw@u.nus.edu}

\author{Sebastian Allende}
\address{Departamento de F\'isica, Universidad de Santiago de Chile (USACH), 
Avenida Ecuador 3493, 9170124, Santiago, Chile}

\author{Leong-Chuan Kwek}
\address{Centre for Quantum Technologies, National University of Singapore,
3 Science Drive 2, Singapore 117543, Singapore}
\address{MajuLab, CNRS-UNS-NUS-NTU International Joint Research Unit, UMI 3654, Singapore}
\address{Institute of Advanced Studies, Nanyang Technological University,
60 Nanyang View, Singapore 639673, Singapore}
\address{National Institute of Education, Nanyang Technological University,
1 Nanyang Walk, Singapore 637616, Singapore}
\author{Guillermo Romero}
\address{Departamento de F\'isica, Universidad de Santiago de Chile (USACH), 
Avenida Ecuador 3493, 9170124, Santiago, Chile}

\begin{abstract}
We study the equilibrium and non-equilibrium physics of two qubits interacting through an ultrastrong coupled qubit-cavity system. By tuning the qubits energy gap while keeping the ultrastrong coupling system to its ground state, we demonstrate a strong two-qubit interaction as well as an enhanced excitation transfer between the two qubits. Our proposal has twofold implications: a means to attain multipurpose parity-protected quantum information tasks in superconducting circuits, and a building block for ultrastrong coupled cavity-enhanced exciton transport in disordered media.
\end{abstract}


\maketitle

\section{Introduction}

Light-matter interaction in the strong coupling regime lies at the root of numerous advances in quantum technologies and  quantum information tasks~\cite{Ladd2010}. Recent experiments in solid state physics have reported an unprecedented coupling between a two-level system (qubit) and an optical/microwave cavity~\cite{Vahala:2003aa}, reaching the ultrastrong (USC)~\cite{PhysRevB.79.201303,Gunter:2009aa,PhysRevLett.105.196402,Scalari1323,PhysRevLett.105.237001,Niemczyk:2010aa,NaturePhys13.39.2017} and deep strong coupling regimes \cite{yoshihara2017}, where the light-matter interaction strength is comparable to or larger than the cavity and qubit frequencies. The USC regime has since been extensively studied in various theoretical contexts~\cite{PhysRevB.72.115303,PhysRevA.74.033811,ANDP:ANDP200710261,PhysRevA.80.032109,PhysRevB.81.235303,PhysRevLett.104.023601,PhysRevB.71.024505,PhysRevLett.107.190402,srep08621,simone_parity,PhysRevLett.114.183601,Rossatto2016}. In the dipolar approximation, the qubit-cavity system can be described by the quantum Rabi model (QRM)~\cite{PhysRev.49.324,PhysRevLett.107.100401}, which features a discrete parity symmetry. Low-lying eigenstates of the QRM hold virtual excitations that cannot escape from the cavity~\cite{PhysRevA.74.033811}, and it has been demonstrated that those eigenstates are useful for parity-protected quantum computing~\cite{PhysRevLett.107.190402} and quantum memory applications~\cite{srep08621}. 

Apart from fundamental interests in light-matter interaction at the USC regime, the ultrastrong coupled qubit-cavity system or quantum Rabi system (QRS) is also extensively studied for its potential impetus to speed up quantum information processing at subnanosecond time scales~\cite{PhysRevB.70.224515,PhysRevB.79.024502,0253-6102-56-1-17,PhysRevLett.108.120501,PhysRevB.91.064503}, particularly within the framework of circuit quantum electrodynamics (QED)~\cite{PhysRevA.69.062320}. For instance, Refs.~\cite{PhysRevLett.108.120501,PhysRevB.91.064503} proposed the use of a tunable qubit-cavity coupling strength~\cite{PhysRevLett.105.023601} to attain ultrafast two-qubit gates~\cite{PhysRevLett.108.120501}. Nevertheless, the major caveat of the above mentioned proposals is the need of various magnetic fluxes acting upon a flux qubit, during quantum gate operations. A typical flux qubit is micrometer in size, thus it is very hard to implement micrometer resolution magnetic field lines threading the qubit without making any interference among them. With our proposed framework presented here, the magnetic crosstalk problem could be overcome, while it might preserve the similar quantum computing performance as Ref.\cite{PhysRevLett.108.120501}. 

Here, we present a parity-preserving USC system that mediates effective two-body interactions, with four compelling characteristics that might have important implications in superconducting circuit-based quantum computing (i-iii) and solid-state physics (iv) communities. (i) Strong two-qubit interaction with an increase in the qubit-cavity coupling strength of the QRS ($g_p/\omega_{\rm cav}$) is demonstrated. (ii) A tunable qubit-qubit interaction could be performed by sweeping only the qubits energy gap for fixed QRS parameters, without requiring complex flux qubit architectures of Refs.~\cite{PhysRevLett.105.023601,PhysRevLett.108.120501}. (iii) Manipulation of the qubits energy gap does not change the underlying $\mathbb{Z}_2$ symmetry, with which generalization to a system, with $N$ qubits and a QRS, can easily be extended; thereby we provide an intuitive physical insight. (iv) Enhanced excitation transfer between the two nonidentical qubits with increase in $g_p/\omega_{\rm cav}$ is shown, while one qubit experiences an incoherent pumping and the other one experiences a loss mechanism. From extensive numerical studies, we will provide an interesting physical insight that might shed some light on the cavity-enhanced exciton transport in disordered medium~\cite{Orgiu:2015aa,PhysRevLett.114.196403,PhysRevLett.114.196402}, especially within the context of polyatomic molecules in the USC regime~\cite{AIP1.4882422}.

This work is organized as follows. In section~\ref{sec:II}, we describe our theoretical model, and discuss its associated parity symmetry. In section.~\ref{sec:III}, we present our results and discussions. Finally, we present our concluding remarks in section~\ref{sec:IV}.   

\begin{figure}[t]
\centering
\includegraphics[scale=0.3]{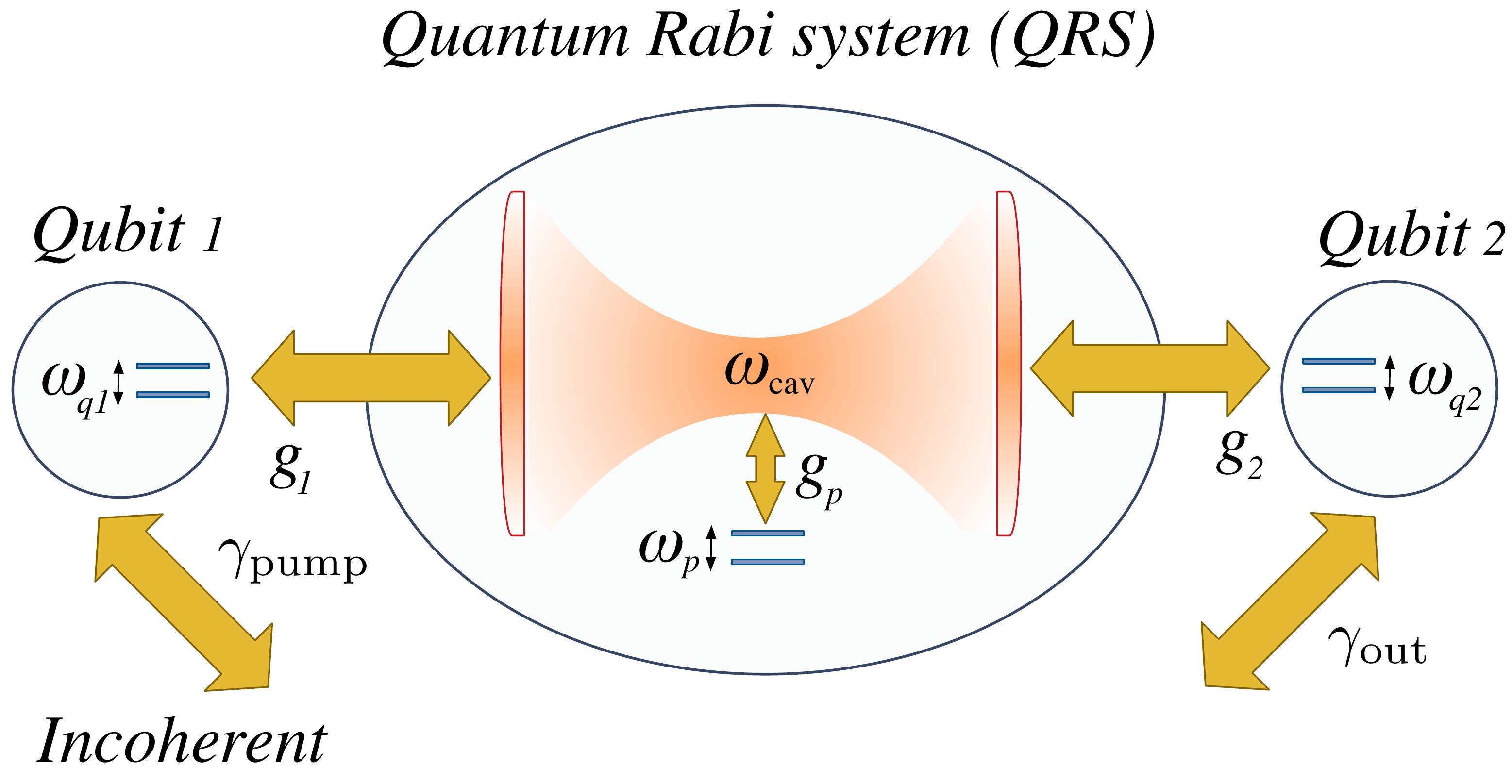}
\caption{Our model schematic. A qubit-cavity system interacting in the USC regime constitutes the QRS, while two additional qubits interact with the cavity mode. This system might also be considered as a building block for exciton transport mechanism, where qubit $1$ is driven by an incoherent pumping, qubit $2$ experiences spontaneous decay, and QRS undergoes lossy channels.}
\label{Fig1}
\end{figure}

\section{The model}
\label{sec:II}
We consider a qubit-cavity system in the USC regime, a QRS (see figure~\ref{Fig1}), which is described by the quantum Rabi model~\cite{PhysRev.49.324,PhysRevLett.107.100401}
\begin{equation}
H_p=\frac{\hbar\omega_p}{2}\sigma^z_p+\hbar\omega_{\rm cav} b^\dag b+\hbar g_p\sigma^x_p(b+b^\dag).
\label{HPolariton}
\end{equation}
Here, $b(b^\dag)$ is the single mode bosonic annihilation(creation) operator and $\sigma^{x,z}$ are the Pauli matrices. We denote $\omega_{\rm cav}$ as the cavity mode frequency, $\omega_p$ and $g_p$ as the qubit frequency and the qubit-cavity coupling strength, respectively. In addition, two qubits with frequencies $\omega_{q1}$ and $\omega_{q2}$ are coupled to the QRS via the cavity mode with coupling strengths $g_1$ and $g_2$. The Hamiltonian for our model reads
\begin{equation}
H = H_p +\sum_{n=1}^{N=2} \frac{\hbar\omega_{qn}}{2}\sigma^z_{n} +\hbar g_n\sigma^x_n(b+b^\dag).
\label{HTotal}
\end{equation} 
This Hamiltonian could be implemented with transmon qubits coupled to a single-mode coplanar waveguide resonator~\cite{PhysRevA.76.042319} (see~\ref{Transmon_Appen}). Notice that our system preserves the following symmetry. If we change $\sigma^x_i \rightarrow-\sigma^x_i$ and $(b+b^\dag) \rightarrow -(b+b^\dag)$, the Hamiltonian in equation~(\ref{HTotal}) remains unchanged, i.e., the system is symmetric with respect to an inversion of the pseudo-spin operator, and the field quadrature. To be precise, one can introduce the parity operator $P=-e^{i\pi b^\dag b}\sigma^{z}_{p}\sigma_1^z\sigma_2^z$ such that $[H,P]=0$. Notice that this parity operator considers the qubit inside the QRS and two additional qubits coupled to the cavity. As the consequence, $H$ and $P$ can be simultaneously diagonalized by $\ket{\phi_j}$, where $P\ket{\phi_j}=\pm \ket{\phi_j}$, and $H\ket{\phi_j}=\hbar \epsilon_j \ket{\phi_j}$, $\forall j$. Also, the parity symmetry establishes selection rules in our system. For instance, states with different parity can only be connected via an interaction that breaks the symmetry, while states with the same parity can only be connected by an interaction that preserves the symmetry. As one varies the qubits energy, this process does not change the parity symmetry, by allowing us to connect states with the same parity as shown in the avoided level crossings of figures~\ref{Fig3}(a) and \ref{Fig4}(a,b).

In the following section, we study the static and dynamical properties of the Hamiltonian~(\ref{HTotal}) as a function of the qubits energy gaps, $\omega_{q1}$ and $\omega_{q2}$, for fixed QRS parameters. This situation could be implemented in circuit QED, where qubits energy level spacing can be tuned via external magnetic fluxes~\cite{Nori:2008aa}. Also, with our proposal, complicated flux qubit architectures~\cite{PhysRevLett.105.023601} are not necessary, thus resolving the magnetic flux crosstalk problem. Moreover, it allows us to achieve an effective qubit-qubit interaction mediated by the QRS, when the latter and two qubits interact dispersively. 

\section{Results and discussions}
\label{sec:III}
In this section, we present the main features of our system. Firstly, we show the equilibrium properties and demonstrate the performance of a tunable strong qubit-qubit interaction in the context of superconducting circuits. Secondly, we show how our system might be used as a primitive unit cell for cavity mediated excitation transfer in the USC regime and its connection with cavity-enhanced exciton transport with organic matter and optical microcavities~\cite{Orgiu:2015aa,PhysRevLett.114.196403,PhysRevLett.114.196402}.   

\subsection{\bf Equilibrium and nonequilibrium properties }
In this subsection, we show that the QRS can be used as a quantum bus mediating an effective two-body interaction. Analogous to the effective interaction mediated by a resonator bus in cavity/circuit QED setups~\cite{PhysRevLett.85.2392,PhysRevA.75.032329,Majer:2007aa}, we demonstrate an effective qubit-qubit interaction as a second-order process, which is a direct consequence of a dispersive coupling between the qubits and QRS. 

To obtain the effective Hamiltonian describing qubit-qubit interaction, we consider a dispersive treatment beyond the rotating wave approximation (RWA). Different from the dispersive theory developed in Ref.~\cite{PhysRevA.80.033846}, we consider the QRS as a whole, and all relevant time scales are compared with all energy level scales, $\omega_{jk}=\omega_j-\omega_k$, where $\omega_j$ satisfy $H_p\ket{j}=\hbar\omega_j\ket{j}$. To proceed, we consider the total Hamiltonian of a system, comprising of $N$ qubits that interact with a QRS, i.e.,
\begin{equation}
	H' = H_0 + H_I,
\end{equation}	
where $H_0=H_p + \sum_{n=1}^N \frac{\hbar \omega_{qn}}{2}\sigma_n ^z$, and the interaction Hamiltonian $H_I = \sum_{n=1}^N\hbar g_n\sigma^x_n(b+b^\dag)$. Onwards, we set $\hbar=1$. If we project the total system onto the QRS eigenbases, we can rewrite $H_p$ as $H_p =\sum_{j=0} ^\infty \omega_j \ketbra{j}{j}$. Using the completeness relation and projecting the interaction Hamiltonian onto the QRS eigenbases, we obtain
\begin{eqnarray}
	H_I &= \sum_{n=1}^N g_n\sigma^x_n \sum_{j,k=0}^\infty \left[\ketbra{k}{k} (b+b^\dag) \ketbra{j}{j} \right]\nonumber\\
	&= \sum_{n,j,k>j} g_n \sigma_n ^x \left[\chi_{kj}\ketbra{k}{j} +\chi_{jk}\ketbra{j}{k}  \right], 
\end{eqnarray}
where $\chi_{kj}=\bra{k} (b+b^\dag) \ket{j}$. Without invoking the RWA, we obtain the interaction Hamiltonian in the interaction picture as
\begin{eqnarray}
	\tilde{H}_I (t) = \sum_{n,j,k>j} g_n &&  \left[ \chi_{jk} e^{i \Delta ^n _{kj}t} \sigma_n ^+ \ketbra{j}{k} + \chi_{kj} e^{i \delta ^n _{kj}t} \sigma_n ^+ \ketbra{k}{j} + \right. \nonumber\\
	&& \left. \chi_{jk} e^{-i \delta ^n _{kj}t} \sigma_n ^- \ketbra{j}{k}+ \chi_{kj} e^{-i \Delta ^n _{kj}t} \sigma_n ^- \ketbra{k}{j} \right].
	\label{Hint}
\end{eqnarray}
Here, $\Delta_{kj}^n =\omega_{qn}-\omega_{kj}$, and $\delta_{kj}^n =\omega_{qn}+\omega_{kj}$. The relevant timescales in our system dynamics come from various energy level differences of the QRS and two qubits frequencies. Here, we are interested in the dispersive limit where the $N$ qubits frequencies are far off-resonant with the lowest QRS transition frequency. In this case, fast oscillatory dynamics can be averaged out to zero and thus only slow dynamics contribute to the overall system dynamics. Hence, we proceed by defining the time average of an operator $\mathcal{O}(t)$ \cite{james2007effective,PhysRevA.82.052106} as 
\begin{equation}
	\overline{\mathcal{O}}(t) \equiv \int_{-\infty} ^\infty f(t-t') \mathcal{O}(t') dt',\label{Heff0}
\end{equation}
where $f(t)$ is real and has unit area. As the consequence, the usual time-ordered evolution operator, satisfying the Schr\"odinger equation
\begin{equation}
	i \frac{\partial}{\partial t} U(t,t_0)=\tilde{H}_I (t) U(t,t_0),
	\label{SEqn}
\end{equation}
can now be rewritten as
\begin{equation}
	i \frac{\partial}{\partial t} \overline{U(t,t_0)}=\mathcal{H}_{\rm eff} (t)\overline{U(t,t_0)},
	\label{Heff1}
\end{equation}
when we invoke the time averaging operator, defined in equation (\ref{Heff0}).
From equations~(\ref{SEqn}-\ref{Heff1}), we find a general expression for $\mathcal{H}_{\rm eff}(t)$ as
\begin{equation}
	\mathcal{H}_{\rm eff}(t) = [\overline{\tilde{H}_I (t) U(t,t_0)}] [\overline{U(t,t_0)}]^{-1},
\end{equation}
where 
\begin{equation}
	U(t,t_0)=\mathcal{T} \exp \left[-i \int_{t_0} ^t \tilde{H}_I (t') dt' \right].
\end{equation}
Here, $\mathcal{T}$ represents a time-ordering operator. Though $U(t,t_0)$ is unitary, its time averaged operator is in general not. Thus, $\mathcal{H}_{\rm eff}(t)$ shown above is not Hermitian, since we have traced out the high-frequency part/s. However, the effective Hamiltonian for the unitary part of the evolution is uniquely given by its Hermitian part \cite{james2007effective}:  
\begin{equation}
	H_{\rm eff}(t)=\frac{1}{2}\{\mathcal{H}_{\rm eff}(t)+\mathcal{H}_{\rm eff}(t)^\dagger \}.	
	\label{Heff2}
\end{equation}
Up to the first order Taylor series expansion, $U(t,t_0)\approx 1+ U_1 (t)$, where $U_1(t) = -i \int_{t_0} ^t dt' \tilde{H}_I (t')$, Eq.~(\ref{Heff2}) can be rewritten as 
\begin{equation}
	{H}_{\rm eff}(t) = \overline{\tilde{H}_I (t)} +\frac{1}{2} \left(\overline{[\tilde{H}_I (t),U_1 (t)]} - [\overline{\tilde{H}_I (t)}, \overline{U_1 (t)}] \right).
	\label{Heff3}
\end{equation}
Explicitly, $U_1 (t)$ has the following expression
\begin{eqnarray}
	U_1 (t) &=& -i  \sum_{n,j,k>j} g_n \left[ \frac{\chi_{jk}}{i\Delta_{kj}^n} \left(e^{i\Delta^n _{kj}t}-1 \right)\sigma_n ^+ \ketbra{j}{k} \right. + \frac{\chi_{kj}}{i\delta_{kj}^n} \left(e^{i\delta^n _{kj}t}-1 \right)\sigma_n ^+ \ketbra{k}{j}\nonumber\\
	 &+&  \frac{\chi_{jk}}{-i\delta_{kj}^n} \left(e^{-i\delta^n _{kj}t}-1 \right)\sigma_n ^- \ketbra{j}{k} + \left. \frac{\chi_{kj}}{-i\Delta_{kj}^n} \left(e^{-i\Delta^n _{kj}t}-1 \right)\sigma_n ^- \ketbra{k}{j}\right].
	 \label{U1}
\end{eqnarray}
By substituting equations~(\ref{Hint}) and (\ref{U1}) into equation~(\ref{Heff3}) and transforming back to the Schr\"odinger picture, we arrive at the following effective Hamiltonian
\begin{eqnarray}
	H_{\rm eff}=& H_0 + \frac{1}{2}\sum_{n,n'}g_ng_{n'}\sum_{j,k>j}\abs{\chi_{jk}}^2\times\nonumber\\
	&\Bigg[\Bigg\{\Bigg(\frac{1}{\Delta^n_{kj}}-\frac{1}{\delta^{n'}_{kj}}\Bigg)\sigma^{+}_{n}\sigma^{+}_{n'}+\Bigg(\frac{1}{\Delta^n_{kj}}+\frac{1}{\Delta^{n'}_{kj}}\Bigg)\sigma^{+}_{n}\sigma^{-}_{n'}\nonumber\\
	&-\Bigg(\frac{1}{\delta^n_{kj}}+\frac{1}{\delta^{n'}_{kj}}\Bigg)\sigma^{-}_{n}\sigma^{+}_{n'}+\Bigg(\frac{1}{\Delta^{n'}_{kj}}-\frac{1}{\delta^{n}_{kj}}\Bigg)\sigma^{-}_{n}\sigma^{-}_{n'}\Bigg\}\ketbra{j}{j}\nonumber\\
	&+\Bigg\{\Bigg(\frac{1}{\delta^n_{kj}}-\frac{1}{\Delta^{n'}_{kj}}\Bigg)\sigma^{+}_{n}\sigma^{+}_{n'}+\Bigg(\frac{1}{\delta^{n'}_{kj}}+\frac{1}{\delta^{n}_{kj}}\Bigg)\sigma^{+}_{n}\sigma^{-}_{n'}\nonumber\\
	&-\Bigg(\frac{1}{\Delta^n_{kj}}+\frac{1}{\Delta^{n'}_{kj}}\Bigg)\sigma^{-}_{n}\sigma^{+}_{n'}+\Bigg(\frac{1}{\delta^{n'}_{kj}}-\frac{1}{\Delta^{n}_{kj}}\Bigg)\sigma^{-}_{n}\sigma^{-}_{n'}\Bigg\}\ketbra{k}{k}\Bigg]
\end{eqnarray}
In deriving the above expression, we have neglected oscillating terms that are proportional to $\exp{(\pm i\omega_{kk'}t)}$. This is provided by the condition $\omega_{kk'}\gg \{g_ng_{n'}|\chi_{jk}|^2 /\Delta^n_{kj},g_ng_{n'}|\chi_{jk}|^2 /\delta^n_{kj}\}$, which can be guaranteed with suitable QRS parameters.  

When we restrict ourselves to the two-qubit case ($N=2$), we arrive at
\begin{eqnarray}
	H_{\rm eff}&=H_0 
	+  \sum_{j,k>j} |\chi_{jk}|^2 \times\nonumber \\
	& \Bigg[ \Bigg\{ \sum_{n=1}^2 \left(\frac{g_n ^2}{\Delta_{kj}^2} \sigma_n ^+ \sigma_n ^- - \frac{g_n ^2}{\delta_{kj}^n} \sigma_n ^- \sigma_n ^+ \right)\nonumber\\
	 &+ \frac{g_1 g_2}{2}\left(\frac{1}{\Delta_{kj}^1}+\frac{1}{\Delta_{kj}^2} -\frac{1}{\delta_{kj}^1}-\frac{1}{\delta_{kj}^2}  \right)\sigma_1 ^x \sigma_2 ^x  \Bigg\} \ketbra{j}{j} \nonumber\\
	&+ \Bigg\{ \sum_{n=1}^2 \left(\frac{g_n ^2}{\delta_{kj}^2} \sigma_n ^+ \sigma_n ^- - \frac{g_n ^2}{\Delta_{kj}^n} \sigma_n ^- \sigma_n ^+ \right)\nonumber\\
	&+ \frac{g_1 g_2}{2}\left(\frac{1}{\delta_{kj}^1}+\frac{1}{\delta_{kj}^2} -\frac{1}{\Delta_{kj}^1}-\frac{1}{\Delta_{kj}^2}  \right)\sigma_1 ^x \sigma_2 ^x  \Bigg\} \ketbra{k}{k} \Bigg].
\end{eqnarray}
Furthermore, when we confine ourselves with the two lowest energy levels of the QRS (i.e., $j=0, k=1$), we obtain
\begin{eqnarray}
	H_{\rm eff}&=&H_0+\frac{1}{2} |\chi_{01}|^2 [\hat{S}_{12}\ketbra{1}{1}-\hat{S}_{12}\ketbra{0}{0}]\nonumber\\
	&=&H_0+\frac{1}{2}|\chi_{01}|^2 \hat{S}_{12}\otimes \hat{Z}_p,\label{Heff}
\end{eqnarray}
where $|\chi_{01}|^2\!\!=\!\!|\bra{0}(b+b^\dagger) \ket{1}|^2$, $\hat{Z}_p\!\!=\!\!\ket{1}\bra{1}-\ket{0}\bra{0}$, $\hat{S}_{12}= g_1 g_2 (1/{\delta^1 _{10}}+1/{\delta^2 _{10}}-1/\Delta^1_{10} -1/{\Delta^2 _{10}})\sigma^x _1 \otimes \sigma^x _2 +\sum_{n=1}^2 2g_n ^2 (\sigma^{+}_n\sigma^{-}_n/{\Delta^n _{10}} -\sigma^{-}_n\sigma^{+}_n/{\delta^n _{10}})$. It is noteworthy that the Schrieffer-Wolff transformation \cite{PhysRev.149.491}, $e^{S}He^{-S}$, applied to the Hamiltonian $(2)$, with the non-Hermitian operator 
\begin{equation}
S=\sum_{n=1}^2\sum_{j,k>j}g_n\Bigg[\frac{\chi_{jk}}{\Delta^n_{kj}}\sigma^+_n\ket{j}\bra{k}-\frac{\chi_{kj}}{\Delta^n_{kj}}\sigma^-_n\ket{k}\bra{j}+\frac{\chi_{kj}}{\delta^n_{kj}}\sigma^+_n\ket{k}\bra{j}-\frac{\chi_{jk}}{\delta^n_{kj}}\sigma^-_n\ket{j}\bra{k}\Bigg],
\end{equation}
and the time-averaging operator method produce, up to a constant term, the same effective qubit-qubit Hamiltonian.

\begin{figure}[t]
\centering
\includegraphics[scale=0.5]{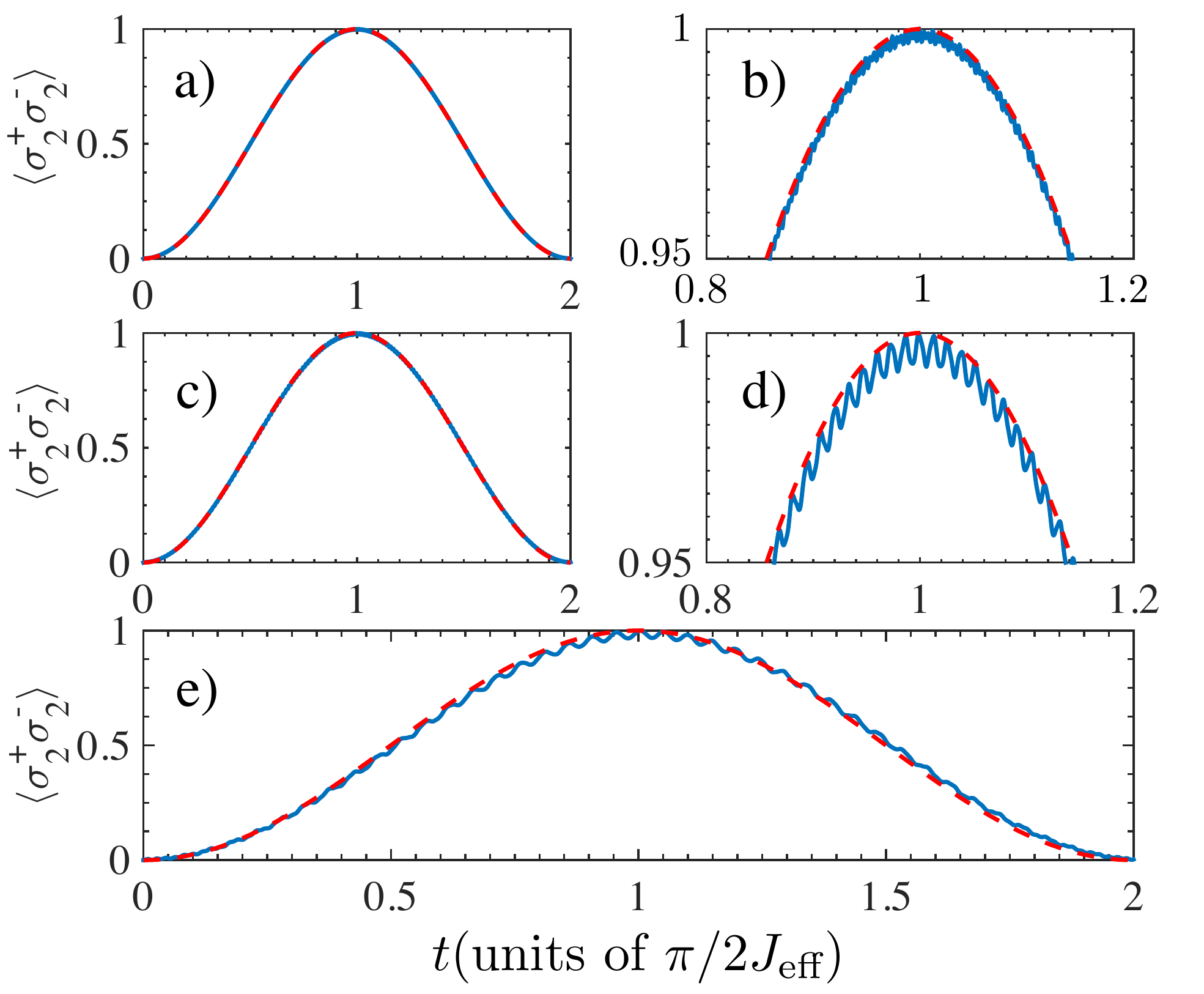}
\caption{Excitation number of the second qubit as a function of time. a) $g_p/\omega_{\rm cav}=0.1$, $\omega_p=0.8\omega_{\rm cav}$, $\omega_{q1}=\omega_{q2}=0.2\omega_{\rm cav}$, and $g_1=g_2=0.02\omega_{\rm cav}$. These parameters lead to an effective qubit-qubit coupling strength $2J_{\rm eff}=0.00176\omega_{\rm cav}$. b) Enlarged portion of (a). c) $g_p/\omega_{\rm cav}=0.3$, $\omega_p=0.8\omega_{\rm cav}$, $\omega_{q1}=\omega_{q2}=0.2\omega_{\rm cav}$, and $g_1=g_2=0.02\omega_{\rm cav}$. These parameters lead to an effective qubit-qubit coupling strength $2J_{\rm eff}=0.00267\omega_{\rm cav}$. d) Enlarged portion of (c). e) $g_p/\omega_{\rm cav}=0.5$, $\omega_p=0.8\omega_{\rm cav}$, $\omega_{q1}=\omega_{q2}=0.2\omega_{\rm cav}$, and $g_1=g_2=0.02\omega_{\rm cav}$. These parameters lead to an effective qubit-qubit coupling strength $2J_{\rm eff}=0.00573\omega_{\rm cav}$. In all the figures, blue (continuous) lines are evolution outcome under the full Hamiltonian, equation~(\ref{HTotal}), and red (dashed) lines are evolution outcome under the effective Hamiltonian, equation~(\ref{Heff}).}
\label{Fig2}
\end{figure}

The first observation we make after going through all this derivation is that the two target qubits interact via the central QRS, since the total system Hamiltonian, equation (\ref{HTotal}), can be approximated by an effective Hamiltonian, equation (\ref{Heff}), when the QRS interacts dispersively with the two qubits. We plot the excitation number of the second qubit $\langle \sigma_2 ^+ \sigma_2 ^- \rangle$ as a function of time in accordance with both Hamiltonians, given that initial system state is at $\ket{0}\otimes\ket{eg}$. The results are shown in figure~\ref{Fig2}, where we use red (dashed) lines for the effective Hamiltonian evolution and blue (continuous) lines for the full one. We see that the two results match pretty well in small $g_p/\omega_{\rm cav}$ parameter regime (cf. figures~\ref{Fig2}(a-d)). However, when $g_p/\omega_{\rm cav}$ is reasonably large as in figure~\ref{Fig2}(e), a clear deviation from the full Hamiltonian dynamics is seen. The reason is that at this coupling strength the QRS gap is closer to the qubits energy splitting such that the QRS can be excited. Moreover, the assumption we make in arriving at the effective Hamiltonian: $\omega_{kk'}\gg \{g_ng_{n'}|\chi_{jk}|^2 /\Delta^n_{kj},g_ng_{n'}|\chi_{jk}|^2 /\delta^n_{kj}\}$ is not true any more. The second observation is that fast oscillations in the full Hamiltonian evolution become apparent with increase in $g_p/\omega_{\rm cav}$, while they are smeared out in the effective Hamiltonian evolutions, because we have employed the time averaging operators of the form in equation (\ref{Heff0}).

\begin{figure*}[t]
\centering
\includegraphics[scale=0.445]{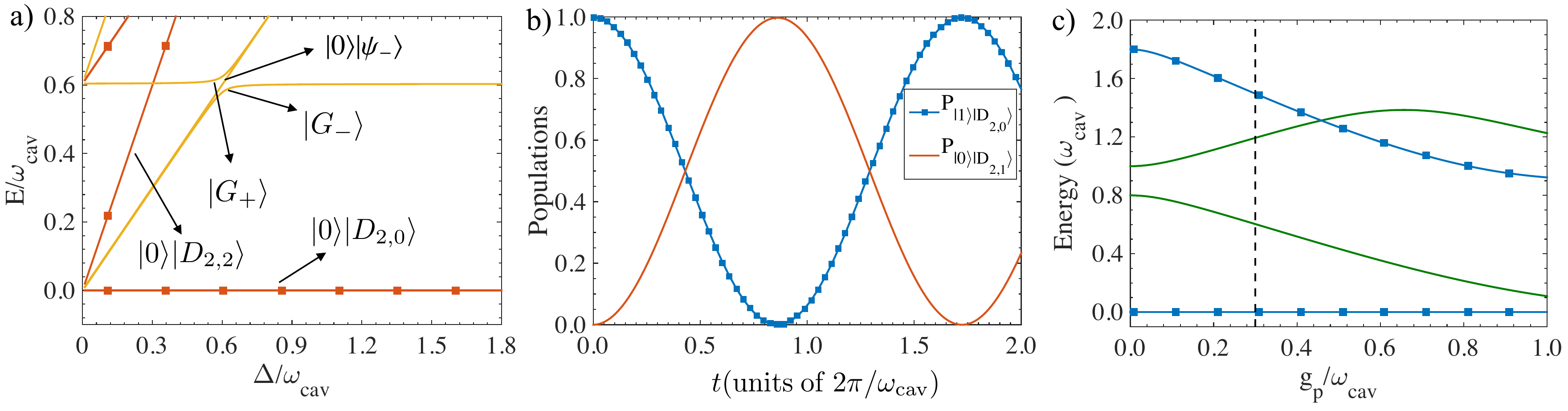}
\caption{a) Energy spectrum ($\hbar=1$) from the Hamiltonian~(\ref{HTotal}) for identical qubits, $\omega_{q1}=\omega_{q2}=\Delta$. Red (squared) lines stand for states with even parity ($p=+1$) and yellow (continuous) lines stand for states with odd parity ($p=-1$), where $p$ is the eigenvalue of the parity operator $P$. b) Population inversion between states $\ket{0}\ket{D_{2,1}}$ and $\ket{1}\ket{D_{2,0}}$ at the resonance condition $\Delta = 0.6042\omega_{\rm cav}$. The numerical calculations for a) and b) are done with parameters $\omega_p=0.8\omega_{\rm cav}$, $g_p=0.3\omega_{\rm cav}$, and $g_1=g_2=0.02\omega_{\rm cav}$. c) Energy spectrum for a single QRS (see equation~(\ref{HPolariton})), with parameters $\omega_p=0.8\omega_{\rm cav}$, as a function of the light-matter coupling. Blue (squared) lines stand for states with even parity and green (continuous) lines for states with odd parity associated with the parity operator ${\Pi}_p=-e^{i\pi b^\dag b}\sigma^{z}_{p}$. The vertical line stands for $g_p/\omega_{\rm cav}=0.3$, corresponding to $\Delta=0$ in (a). The states indicated here are not the actual eigenstates of the total Hamiltonian, equation (\ref{HTotal}). They are approximate states that are calculated in the truncated subspace $\{\ket{0},\ket{1}\}\otimes\{\ket{gg},\ket{ge},\ket{eg},\ket{ee}\}$.}
\label{Fig3}
\end{figure*}

Now, we are in the position to discuss the qubit-qubit interaction. We first consider the simplest scenario, where two identical qubits interact with a QRS. In this case, the Hilbert space of the whole system is spanned by tensor products of the QRS eigenbasis $\{|j\rangle\}$ ($j=0,1,...,\infty$), and the symmetric Dicke states $\{|D_{N,k}\rangle\}$ with $N=2$ qubits and $k=0,1,2$ excitation number. Namely, $|D_{2,0}\rangle=|gg\rangle$, $|D_{2,1}\rangle=(|eg\rangle+|ge\rangle)/\sqrt{2}$, and $|D_{2,2}\rangle=|ee\rangle$. Figure~\ref{Fig3}(a) shows the lowest energy states as a function of the qubits energy gap $\omega_{q1}\!=\!\omega_{q2}\!=\!\Delta$. Notice that varying the qubits energy gap does not change the underlying ${\mathbb{Z}}_2$ symmetry. As the energy gap approaches the QRS energy, precisely at $\Delta = 0.6042\omega_{\rm cav}$, the energy spectrum shows avoided level crossing between the states with the same parity. In addition, the spectrum shows a straight line representing the state $\ket{0}\ket{\psi_{-}}$, where $\ket{\psi_{-}}=(\ket{eg}-\ket{ge})/\sqrt{2}$ is the singlet state that does not couple with the QRS. However, this state does not appear in spectroscopic measurements~\cite{Bretheau:2013aa,Janvier1199}. 
The lowest energy states shown in figure~\ref{Fig3}(a) are linear superposition of the QRS and two qubits states $\{|j\rangle|D_{N,k}\rangle\}$. At the first avoided level crossing with the resonance condition $\Delta=0.6042\omega_{\rm cav}$, both the qubits and QRS are maximally entangled. They can be approximated by $\ket{G_\pm}\approx\ket{0}\ket{D_{2,1}}\pm\ket{1}\ket{D_{2,0}})/\sqrt{2}$. In particular, states $\ket{0}\ket{D_{2,1}}$ and $\ket{1}\ket{D_{2,0}}$ exhibit population inversion as shown in figure~\ref{Fig3}(b). Notice that the QRS energy spectrum is recovered at $\Delta=0$ in figure~\ref{Fig3}(a) (see figure.~\ref{Fig3}(c), where the dashed vertical line depicts the $\Delta=0$ case).  We note that the lowest states that appear in figure~\ref{Fig3}(a) are approximated states of the total Hamiltonian equation (\ref{HTotal}), when the latter is truncated to the basis defined by the states $\{\ket{0},\ket{1}\}\otimes\{\ket{gg},\ket{ge},\ket{eg},\ket{ee}\}$.

\begin{figure*}[t]
\centering
\includegraphics[scale=0.165]{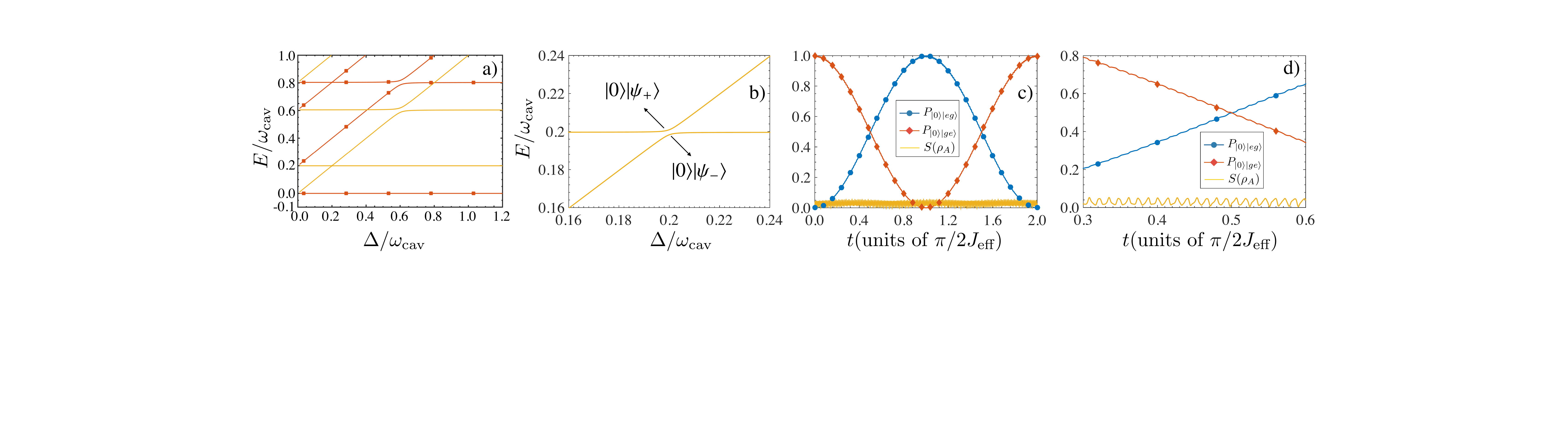}
\caption{a) Energy spectrum ($\hbar=1$) of the Hamiltonian Eq.~(\ref{HTotal}) as a function of qubit $2$ energy, $\omega_{q2}=\Delta$. Here, we consider parameters $\omega_p=0.8\omega_{\rm cav}$, $g_p=0.3\omega_{\rm cav}$, $\omega_{q1}=0.2\omega_{\rm cav}$, and $g_1=g_2=0.02\omega_{\rm cav}$. Red (squared) lines stand for states with even parity and yellow (continuous) lines stand for states with odd parity. b) First avoided level crossing at $\Delta=0.2\omega_{\rm cav}$ is enlarged. c) Population inversion between states $\ket{0}\ket{eg}$, blue (circle) line, and $\ket{0}\ket{ge}$, red (diamond) line, at the first avoided level crossing is shown. The yellow (continuous) line shows the Von Neumann entropy between the QRS and two qubits, indicating negligible entanglement between the two subsystems. d) We enlarge fast oscillations seen in c). The states indicated in a) and b) are not the actual eigenstates of the total Hamiltonian, equation (\ref{HTotal}). They are approximate states in dispersive limit.}
\label{Fig4}
\end{figure*}

Until now, all the numerical analyses are done with the assumption that the two qubits are identical. One natural scenario is when the two qubits are not identical, which is always the case for superconducting qubits. In figure \ref{Fig4}, we report numerical study with non-identical qubits. Similar to the identical qubits case reported in figure \ref{Fig3}, we also observe an avoided level crossing at $\Delta/\omega_{\rm cav}=0.2$, without the presence of the non-interacting state $\ket{0}\ket{\psi_-}$, represented by a straight line in figure \ref{Fig3}(a). Figure~\ref{Fig4}(a) shows the energy spectrum of the total system, comprising of the QRS and two qubits, while qubit $1$ energy gap is fixed at $\omega_{q1}=0.2 \omega_{\rm{cav}}$ and qubit $2$ energy gap is varied from $\omega_{q2} \in [0,1.2] \omega_{\rm cav}$. The enlarged diagram of figure \ref{Fig4}(a) is shown in figure~\ref{Fig4}(b), where we identify the two states approximately as $\ket{0}\ket{\psi_{+}}$ and $\ket{0}\ket{\psi_{-}}$, where $\ket{\psi_{\pm}}=(\ket{eg}\pm\ket{ge})/\sqrt{2}$. At the crossing, these two states have an energy gap $\Delta_{\rm gap}=2J_{\rm eff}$. To show this is the case, we plot the populations $P_{\ket{0}\ket{eg}}$ and $P_{\ket{0}\ket{ge}}$ as a function of time with a prior initial state $\ket{\bar{\psi}}=\ket{0}\ket{eg}$. The result is shown in figure~\ref{Fig4}(c). The von Neumann entropy, $S(\rho_A)$, between the QRS and two qubits is also plotted with yellow solid line in the same figure. We observe negligible entanglement between the two subsystems, which guarantees that the QRS is not excited during the qubit-qubit interaction. In addition, we see fast oscillations in the population shown in figure \ref{Fig4}(c,d). One can show that these fast oscillations have a negligible contribution for the weak qubit-QRS coupling strength, i.e., $g_1, g_2 \ll g_p$, and the dynamics predicted from the full Hamiltonian (\ref{HTotal}) and the effective Hamiltonian, equation~(\ref{Heff}), match perfectly well, as we have discussed earlier on when reporting the results in figure \ref{Fig2}.

In a realistic implementation for observing Rabi oscillations between states $\ket{eg}$ and $\ket{ge}$, one possibility is to make use of Gaussian and Stark control pulses as described in Ref. \cite{Majer:2007aa}, where a strong qubit-qubit interaction is mediated via a cavity bus. The protocol for entangling both target qubits works as follows. Step 1: we let the system to cool down to its ground state $\ket{0}\ket{gg}$. Step 2: we apply a Gaussian $\pi$ pulse acting upon the target qubit 1 in order to prepare the state $\ket{0}\ket{eg}$. Step 3: a Stark pulse is applied to the target qubit 2 bringing the qubits into resonance for a variable time $\Delta t$. In this way, the strong qubit-qubit interaction will induce the desired population transfer between states $\ket{eg}$ and $\ket{ge}$.

In our proposal, the time for entangling two qubits is drastically reduced with increasing $g_p/\omega_{\rm cav}$. For instance, if we consider the cavity frequency $\omega_{\rm cav}\!=\!2\pi\times 8~\rm{GHz}$ for the USC system as in Ref.~\cite{PhysRevLett.105.237001}, our model predicts effective qubit-qubit coupling strengths of $2J_{\rm eff}\!=\!2\pi\times (21,28,46)~{\rm MHz}$ for $g_p/\omega_{\rm cav}=(0.3,0.4,0.5)$, respectively. These values lead to times of $t=\!\pi/(4J_{\rm eff})\!\approx\! (11,9,5)~{\rm ns}$, which scale similar to or better than the time needed to perform a controlled phase-gate with fast and resonant gates in new-generation circuit QED setups~\cite{PhysRevB.82.024514,PhysRevLett.112.070502,PhysRevLett.113.220502}. Indeed, one could increase the effective qubit-qubit coupling strength by increasing the ratio $g_p/\omega_{\rm cav}$ in the QRS. For instance, with coupling strengths of $g_p/\omega_{\rm cav}=(0.6,0.8)$, and cavity frequency of $\omega_{\rm cav}=2\pi\times 8~{\rm GHz}$, one can reach effective coupling strengths of about $2J_{\rm eff}= 2\pi\times(77,160)~{\rm MHz}$. These values lead to times of $t=\pi/(4J_{\rm eff})\approx(3,1.5)~{\rm ns}$. However, the latter coupling parameters $g_p/\omega_{\rm cav}$ violate the dispersive interaction between the QRS and the two qubits, since the QRS gap closes as $g_p/\omega_{\rm cav}$ increases. A single restriction of our proposal is to work within a parameter range for $g_p/\omega_{\rm cav}$ that enables the dispersive interaction. We stress that our scheme does not require a tunable qubit-cavity coupling as in the case of circuit QED-based ultrafast gates~\cite{PhysRevLett.105.023601,PhysRevLett.108.120501}. These results establish the features (i-iii) outlined in ``Introduction", and underline one of the main attributes of the paper.
\begin{figure}[t]
\centering
\includegraphics[scale=0.5]{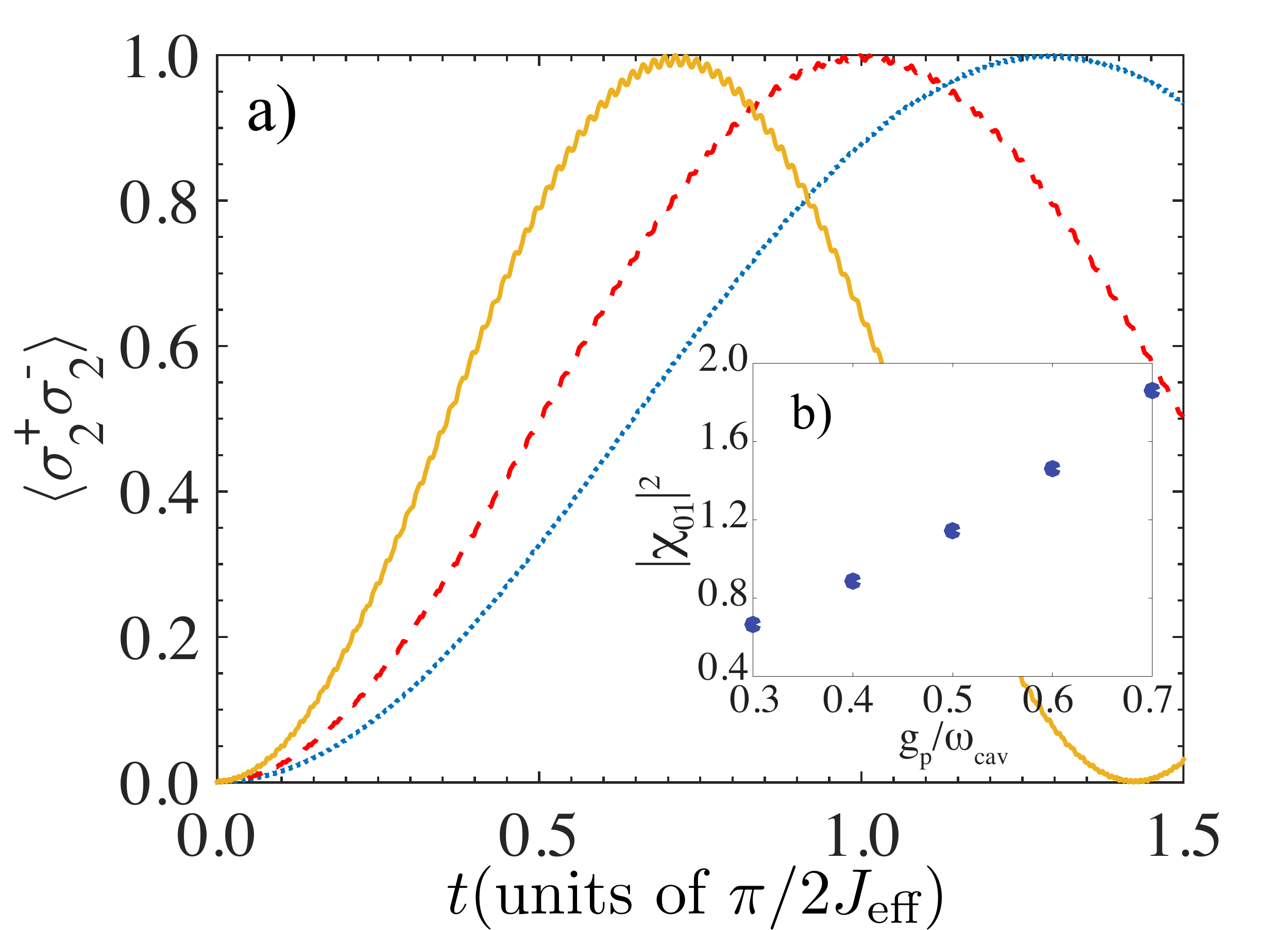}
\caption{a) Excitation number of the qubit $2$ as a function of time. Yellow (continuous) line stands for $g_p=0.4\omega_{\rm{cav}}$, red (dashed) line stands for $g_p=0.3\omega_{\rm{cav}}$, and blue (dotted) line stands for $g_p=0.2\omega_{\rm{cav}}$. In all these cases, we consider $2J_{\rm eff}=0.00267\omega_{\rm cav}$ corresponding to parameters $\omega_p=0.8\omega_{\rm cav}$, $g_p=0.3\omega_{\rm{cav}}$, $\omega_{q1}=\omega_{q2}=0.2\omega_{\rm cav}$, and $g_1=g_2=0.02\omega_{\rm cav}$. b) (inset) The matrix element $|\chi_{01}|^2$ is plotted against $g_p/\omega_{\rm cav}$.}
\label{Fig5}
\end{figure}

Improvement in excitation transfer is achieved with increase in $g_p/\omega_{\rm cav}$ as shown in figure~\ref{Fig5}, where we study the excitation number of the qubit $2$ as a function of $g_p/\omega_{\rm cav}$, with the initial input state $\ket{0}\ket{eg}$. We observe that the larger $g_p/\omega_{\rm cav}$, the faster the rate of excitation transfer. There are two physical interpretations. The first one is that an increase in the matrix element $|\chi_{01}|^2$ with increasing $g_p/\omega_{\rm cav}$, as seen in figure~\ref{Fig5}(b). The second one is that the hybridized cavity frequency is reduced, thus reducing the energy cost of the virtual process mediated by the quantum Rabi system. We note that similar improvement of excitation transfer in a linear resonator bus~\cite{PhysRevLett.85.2392,PhysRevA.75.032329,Majer:2007aa}, unlike non-linear QRS bus discussed here, can in principle be attainable by changing the resonator frequency or its impedance. 

\subsection{\bf Excitation transfer between qubits} Our next task is motivated by recent developments in cavity-enhanced energy and charge transport with organic matter in optical microcavities \cite{Orgiu:2015aa,PhysRevLett.114.196403,PhysRevLett.114.196402,AIP1.4882422} in the strong coupling regime, in conjunction with the results obtained in figure \ref{Fig5}. As seen in the figure, we see increase in the qubit-qubit interaction as we increase $g_p /\omega_{\rm cav}$. The natural question from this simple observation is then how the total system behaves in the long run. Therefore, we proceed to investigate steady-state excitation transfer towards a spontaneously emitting acceptor (modelled with qubit $2$), when a donor (qubit $1$) is incoherently pumped at rate $\gamma_{\rm pump}$. Both the donor and acceptor are mediated by a lossy QRS. 

Our idea follows from the photovoltaic models of Refs. \cite{C6CP06098F,Dorfman19022013}, where energy is assumed to dissipate in a reaction center at the final stage of the energy transfer chain. The reaction center physically corresponds to an acceptor molecule that undergoes a charge separation event, thus producing electric work. The extent of charge separation in a reaction center is proportional to the population of acceptor excited state, which is the observable we monitor in our numerical calculations. We stress that our case of study represents a minimal model for a photovoltaic cell that needs to be further explored in order to propose a realistic photovoltaic array.   

Since our system governed by equation~(\ref{HTotal}) has a large anharmonicity, whose eigenvalues and eigenvectors are defined by $H\ket{\phi_j}=\epsilon_j\ket{\phi_j}$, we need to consider the coloured nature of baths and the hybridization of the qubit-cavity operators \cite{PhysRevA.80.053810,PhysRevA.84.043832,PhysRevLett.109.193602,PhysRevB.89.104516,PhysRevApplied.4.054001}. Under these considerations, we follow the microscopic master equation described in Refs. \cite{PhysRevA.84.043832,PhysRevLett.109.193602}, which reads
\begin{equation}
\dot{\rho}(t)=-i[H,\rho(t)]+\sum_{n=x_1,x_2,x_p,z_p,b}\mathcal{L}_{n}\rho(t).
\label{Master1}
\end{equation} 
Here $\mathcal{L}_{x_1}$ and $\mathcal{L}_{x_2}$ are Liouvillian superoperators describing the incoherent pumping upon qubit $1$ ($\gamma_{\rm pump}$) and the spontaneous emission of qubit $2$ ($\gamma_{\rm out}$). Moreover, we include loss mechanisms acting on the QRS via transversal noise ($\gamma_x$), longitudinal noise ($\gamma_z$), and noise acting on the field quadrature ($\gamma_{\rm cav}$), through Liouvillian superoperators $\mathcal{L}_{x_p}$, $\mathcal{L}_{z_p}$, and $\mathcal{L}_{b}$. In particular, $\mathcal{L}_{x_1}\rho(t)=\sum_{j,k>j}\Gamma^{jk}_{x_1}\mathcal{D}[\ket{\phi_k}\bra{\phi_j}]\rho(t)$ and $\mathcal{L}_{\sigma}\rho(t)=\sum_{j,k>j}\Gamma^{jk}_{\sigma}\mathcal{D}[\ket{\phi_j}\bra{\phi_k}]\rho(t)$ for $\sigma=x_2,x_p,z_p, b$, where $\mathcal{D}[\mathcal{O}]\rho(t)=\frac{1}{2}[2\mathcal{O}\rho(t)\mathcal{O^{\dag}}-\rho(t)\mathcal{O^{\dag}}\mathcal{O}-\mathcal{O^{\dag}}\mathcal{O}\rho(t)]$. The frequency dependent rates are $\Gamma^{jk}_{x_1}=\gamma_{\rm pump}\frac{\epsilon_{kj}}{\omega_{q1}}|\bra{\phi_j}\sigma^x_1\ket{\phi_k}|^2$, $\Gamma^{jk}_{x_2}=\gamma_{\rm out}\frac{\epsilon_{kj}}{\omega_{q2}}|\bra{\phi_j}\sigma^x_2\ket{\phi_k}|^2$, $\Gamma^{jk}_{x_p}=\gamma_{x}\frac{\epsilon_{kj}}{\omega_{p}}|\bra{\phi_j}\sigma^x_p\ket{\phi_k}|^2$, $\Gamma^{jk}_{z_p}=\gamma_{z}\frac{\epsilon_{kj}}{\omega_{p}}|\bra{\phi_j}\sigma^z_p\ket{\phi_k}|^2$, and $\Gamma^{jk}_{b}=\gamma_{\rm cav}\frac{\epsilon_{kj}}{\omega_{\rm cav}}|\bra{\phi_j}(b+b^{\dag})\ket{\phi_k}|^2$, where $\epsilon_{kj}=\epsilon_k-\epsilon_j$. 

The steady state solution of the density matrix can be found by means of the superspace operator method described in Ref.~\cite{navarrete2015open}. If the system of interest belongs to a Hilbert space of dimension ${\rm dim}({\mathcal{H}})=d$, the master equation in the superspace method reads
\begin{eqnarray}
\ket{\dot{\rho}}&=&\left[-i~H\otimes \mathbb{I}+i~\mathbb{I}\otimes H^{T} + \sum_{n=x_1,x_2,x_p,z_p,b}\tilde{\mathcal{L}_n}\right]\ket{\rho},
\end{eqnarray}
where each superoperator $\tilde{\mathcal{L}_n}\propto \frac{1}{2}[2\mathcal{O}\otimes\mathcal{O}^*-\mathbb{I}\otimes\mathcal{O}^T\mathcal{O}^*-\mathcal{O}^\dag\mathcal{O}\otimes\mathbb{I}]$. Notice that the superspace dimension is ${\rm dim}(\mathcal{S})=d^2$, and $\ket{\dot{\rho}}=d\ket{\rho}/dt$.  In the master equation above, $\mathbb{I}$ is the superspace identity and we take into account the transpose and conjugate of system operators. The steady state solution ($\rho_{\rm ss}$) is found numerically under the condition $\ket{\dot{\rho}}=0$. This implies to find the eigenstate of $\mathcal{L}=-i~H\otimes \mathbb{I}+i~\mathbb{I}\otimes H^{T} + \sum_{n}\tilde{\mathcal{L}_n}$, with zero eigenvalue, i.e., $\mathcal{L}~\ket{\rho}_{\rm ss}=0$.

\begin{figure}[t]
\centering
\includegraphics[scale=0.5]{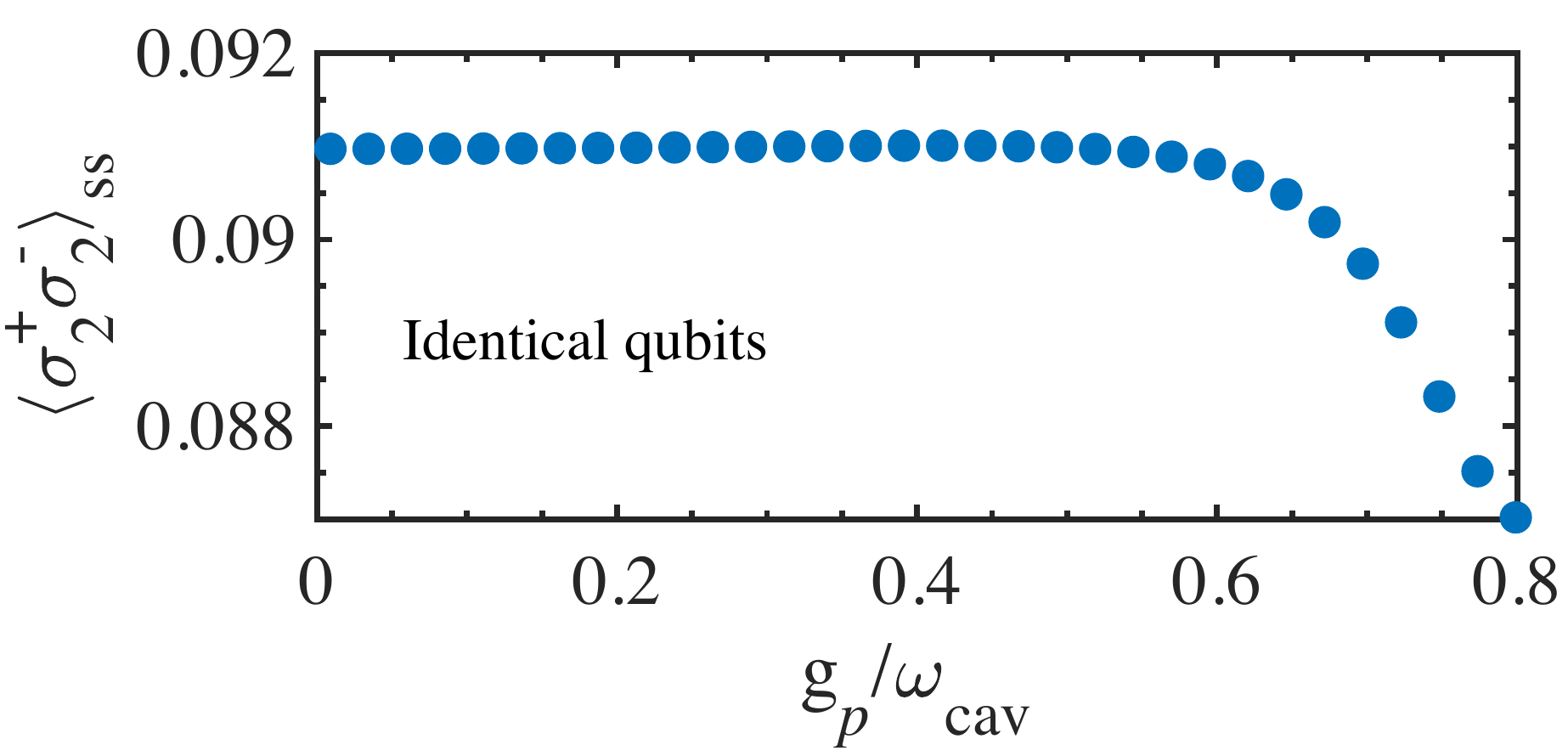}
\caption{Steady-state excitation number of qubit $2$ versus $g_p/\omega_{\rm cav}$. In this simulation, $\omega_{q1}=\omega_{q2}=0.2\omega_{\rm cav}$, $\omega_p=0.8\omega_{\rm cav}$, $g_1=g_2=10^{-2}\omega_{\rm cav}$ are considered. The pumping and loss rates are $\gamma_{\rm pump}=\gamma_x=\gamma_z=\gamma_{\rm cav}=10^{-2}\omega_{\rm cav}$, and $\gamma_{\rm out}=10^{-1}\omega_{\rm cav}$. We note that a Fock space of $N=8$ is enough to assure convergence for each value of the ratio $g_p/\omega_{\rm cav}$.}
\label{Fig6}
\end{figure}

We study the steady-state excitation transferred to qubit $2$ ($\langle\sigma^{+}_2\sigma^{-}_2\rangle_{\rm ss}$) for two cases, namely, identical qubit case and nonidentical one. For the identical qubit case, the steady state population of the qubit $2$, obtained from the master equation (\ref{Master1}) with the {\it ab initio} Hamiltonian (\ref{HTotal}), exhibits a flat behavior until $g/\omega_{\rm cav}\approx 0.6$ as shown in figure \ref{Fig6}.
For the chosen parameters mentioned in figure \ref{Fig6}, the value $g/\omega_{\rm cav}\approx 0.6$ establishes a limit where both qubits departs from the dispersive regime with the QRS. Beyond this point, physics is not captured by virtual excitation of the QRS, since the latter and the qubits become resonance. 

\begin{figure}[t]
\centering
\includegraphics[scale=0.5]{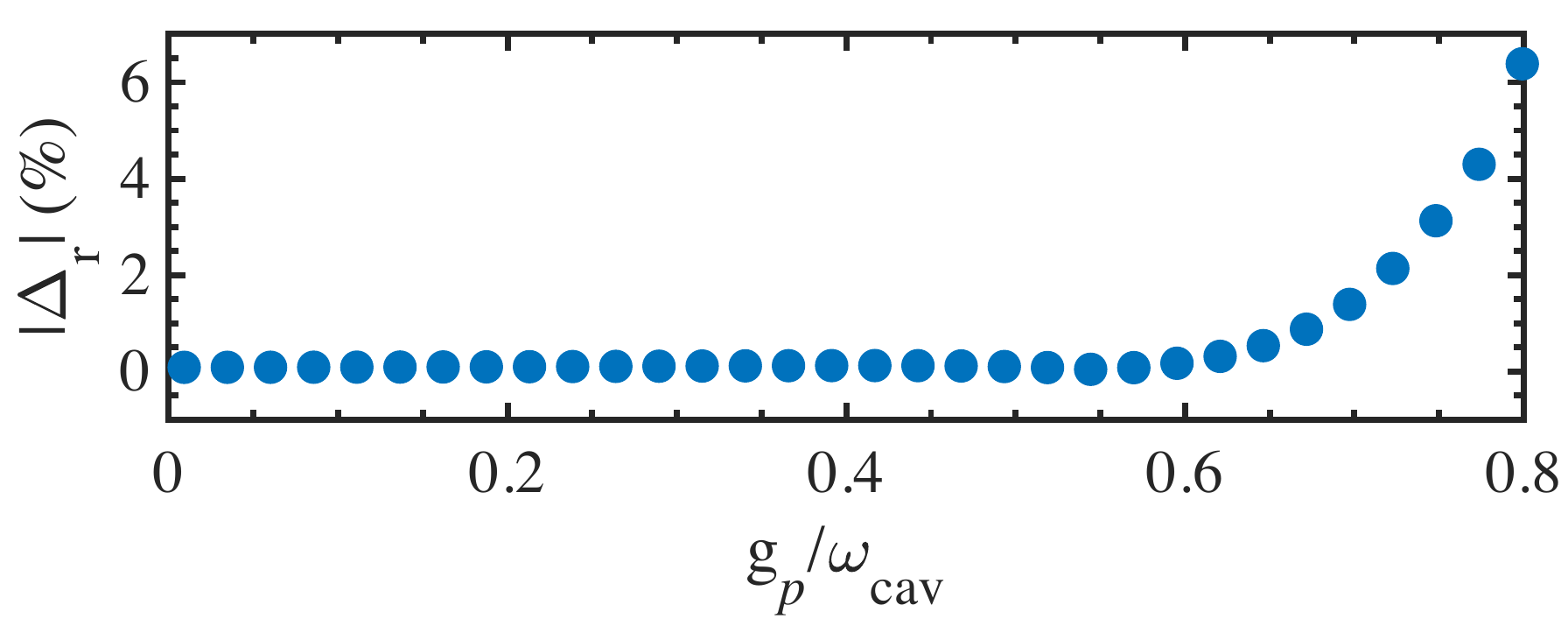}
\caption{The absolute value of the relative difference $\Delta_{\rm r}=1-\langle\sigma^{+}_2\sigma^{-}_2\rangle^{\rm effective}_{\rm ss}/\langle\sigma^{+}_2\sigma^{-}_2\rangle^{\it ab~initio}_{\rm ss}$ in percentage, computed numerically from the {\it ab initio} Hamiltonian (\ref{HTotal}) ($\langle\sigma^{+}_2\sigma^{-}_2\rangle^{\it ab~initio}_{\rm ss}$) and from the effective two-qubit Hamiltonian (\ref{HeffA}) ($\langle\sigma^{+}_2\sigma^{-}_2\rangle^{\rm effective}_{\rm ss}$), is plotted against $g_p/\omega_{\rm cav}$, with the same parameters of figure \ref{Fig6}.}
\label{Fig7}
\end{figure}

The flat behavior when decreasing the $g/\omega_{\rm cav}$ has an intuitive explanation if we study the system eigenstates of the effective qubit-qubit Hamiltonian (\ref{Heff}). There, the effective qubit-qubit coupling strength $J_{\rm eff}$ is very small compared with the qubits frequencies $\omega_{q1}$ and $\omega_{q2}$ such that one can perform the rotating wave approximation in the qubit-qubit interaction
\begin{equation}
H_{\rm eff} \approx \sum^2_{n=1}\frac{\tilde{\omega}_{qn}}{2}\sigma^{z}_n + J_{\rm eff}(\sigma^+_1\sigma^-_2+\sigma^-_1\sigma^+_2),
\label{HeffA}
\end{equation}
where $\tilde{\omega}_{qn}=\omega_{qn}+|\chi_{01}|^2g^2_n(1/\Delta^n_{10}+1/\delta^n_{10})$ and $J_{\rm eff}=\frac{1}{2}|\chi_{01}|^2g_1g_2(1/\delta^1_{10}+1/\delta^2_{10}-1/\Delta^1_{10}-1/\Delta^2_{10})$. The eigenvalues and eigenstates of Hamiltonian (\ref{HeffA}) are $E_G=-\frac{1}{2}\sqrt{\tilde{\omega}_{q1}+\tilde{\omega}_{q2}}$, $E_1=-\frac{1}{2}\sqrt{4J^2_{\rm eff}+(\tilde{\omega}_{q1}-\tilde{\omega}_{q2})^2}$, $E_2=\frac{1}{2}\sqrt{4J^2_{\rm eff}+(\tilde{\omega}_{q1}-\tilde{\omega}_{q2})^2}$, $E_3=\frac{1}{2}\sqrt{\tilde{\omega}_{q1}+\tilde{\omega}_{q2}}$ and $\ket{G}=\ket{gg}$, $\ket{E_1}=-\sin(\theta/2)\ket{eg}+\cos(\theta/2)\ket{ge}$, $\ket{E_2}=\cos(\theta/2)\ket{eg}+\sin(\theta/2)\ket{ge}$, $\ket{E_3}=\ket{ee}$, with $\tan(\theta)=2J_{\rm eff}/(\tilde{\omega}_{q1}-\tilde{\omega}_{q2})$. It is apparent that for identical qubit frequencies $\omega_{q1}=\omega_{q2}$ and qubit-QRS coupling strengths $g_1=g_2$, the eigenstates of the joint qubit-qubit system do not depend on the effective coupling strength $J_{\rm eff}$. This implies that neither the matrix elements of operators appearing in the microscopic master equation for the effective two-qubit system
\begin{equation}
\dot{\rho}_Q(t)=-i[H_{\rm eff},\rho_Q(t)]+\sum_{n=x_1,x_2}\mathcal{L}_{n}\rho_Q(t),
\label{Master2}
\end{equation} 
where $\rho_Q$ describes the two qubits density matrix, nor the energy differences will have an influence on $J_{\rm eff}$, since the dominating frequency scale is $\omega_{q1}$ ($\omega_{q2})$. Notice that the Liouvillian superoperator $\mathcal{L}_n$ needs to be evaluated in terms of the effective qubit-qubit basis described above. We perform numerical calculations to obtain the steady state solution $\langle\sigma^{+}_2\sigma^{-}_2\rangle_{\rm ss}$ computed numerically from the master equation (\ref{Master1}) with the {\it ab initio} Hamiltonian (\ref{HTotal}) ($\langle\sigma^{+}_2\sigma^{-}_2\rangle^{\it ab~initio}_{\rm ss}$) and the one from the master equation (\ref{Master2}) with the effective two-qubit Hamiltonian (\ref{HeffA}) ($\langle\sigma^{+}_2\sigma^{-}_2\rangle^{\rm effective}_{\rm ss}$). Figure \ref{Fig7} shows the absolute value of the relative difference $\Delta_{\rm r}=1-\langle\sigma^{+}_2\sigma^{-}_2\rangle^{\rm effective}_{\rm ss}/\langle\sigma^{+}_2\sigma^{-}_2\rangle^{\it ab~initio}_{\rm ss}$ in percentage, for the same parameters used in figure \ref{Fig6}. 

\begin{figure}[t]
\centering
\includegraphics[scale=0.5]{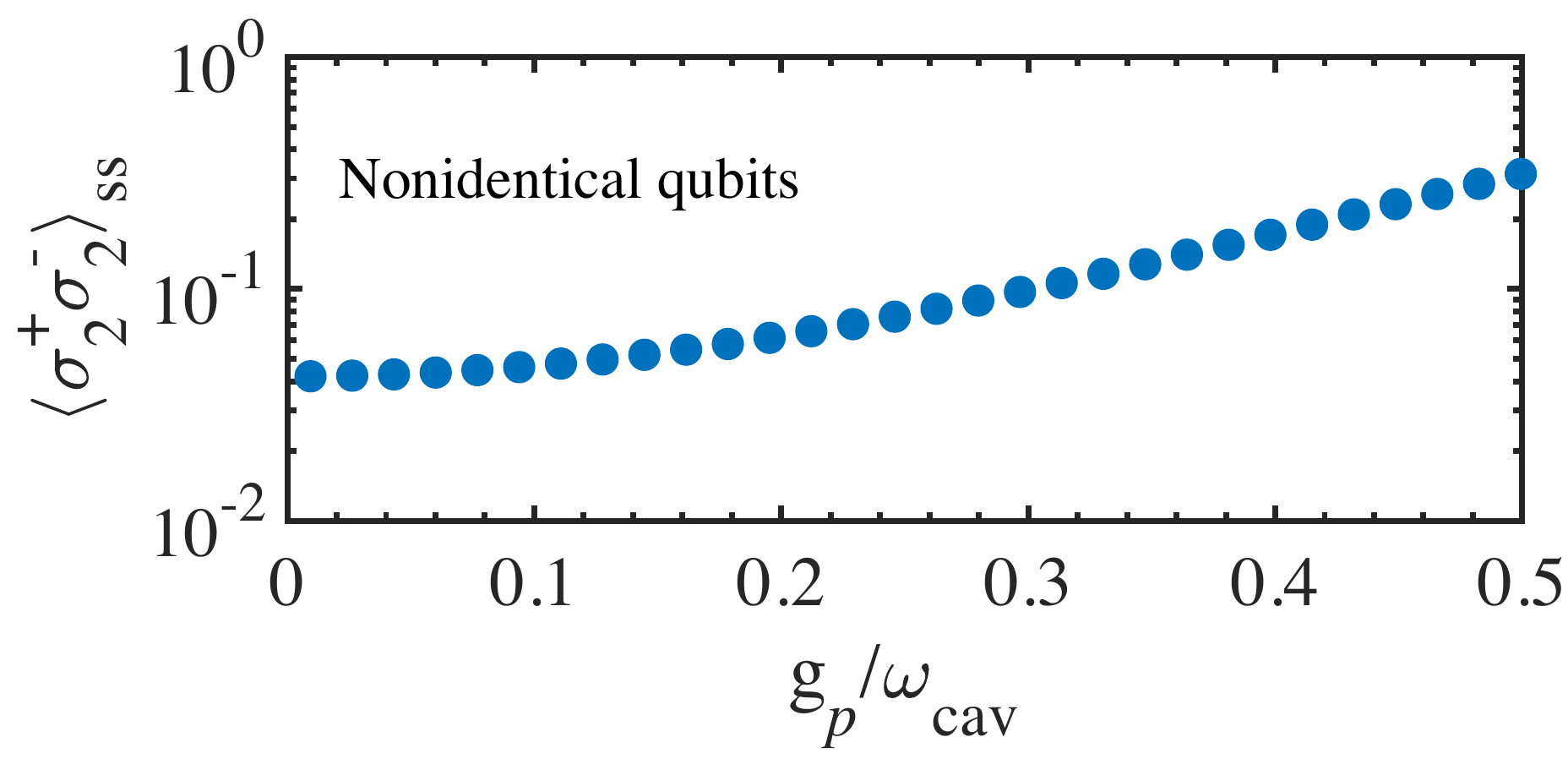}
\caption{Steady-state excitation number of qubit $2$ versus $g_p/\omega_{\rm cav}$, for the case of nonidentical qubits. In this simulation we used parameters $\omega_{q1} = 0.2\omega_{\rm cav}$, $\omega_{q2} = 0.19\omega_{\rm cav}$, $\omega_p = 0.8\omega_{\rm cav}$, $g_1 = g_2 = 10^{-2}\omega_{\rm cav}$, $\gamma_{\rm out} = 10^{-4}\omega_{\rm cav}$, $\gamma_{\rm pump} = \gamma_x = \gamma_z = \gamma_{\rm cav}=10^{-2}\omega_{\rm cav}$.}
\label{Fig8}
\end{figure}

For nonidentical qubit case where $\omega_{q1}\neq\omega_{q2}$, the result of the steady state mean value of the excitation number of qubit $2$, obtained from the master equation (\ref{Master1}) with the {\it ab initio} Hamiltonian (\ref{HTotal}), is shown in Figure \ref{Fig8}. As the coupling strength of the QRS enters the USC regime, $0.1\lesssim g_p/\omega_{\rm cav}<1$, we see a striking one order of magnitude increase in the excitation transfer as $g_p/\omega_{\rm cav}$ increases from $10^{-2}$ (the strong coupling regime) to $0.5$ (the USC regime). It is noteworthy that the nonidentical qubits case represents a more realistic approach towards donor-acceptor organic photovoltaic complex, where donor and acceptor molecules are not identical. This result establishes the feature (iv).

\section{Conclusions}
\label{sec:IV}
In this paper, we have investigated the equilibrium and non-equilibrium dynamics of two qubits interacting via a QRS. We have demonstrated a possibility of using the QRS as a quantum bus to mediate a strong and tunable qubit-qubit interaction. In particular, two-qubit entanglement time can be reduced as qubit-cavity coupling in the QRS increases. We also highlight that the manipulation of the qubits energy relaxes the requirement of complex flux qubit architectures, and thereby without requiring tunable qubit-cavity coupling to perform qubit-qubit interaction as in Ref. \cite{PhysRevLett.108.120501}. Hence, our theoretical proposal could be implemented in a circuit QED setup with existing technologies. In particular, one can consider a flux qubit coupled galvanically to an on-chip $\lambda/2$ transmission line resonator to form the  QRS~\cite{Niemczyk:2010aa}. Two additional transmon qubits are positioned at the resonator edges, where the resonator voltage is maximum. In this way, possible magnetic crosstalk between on-chip flux lines can be avoided. Furthermore, since transmons are many-level systems rather than qubits, additional analysis is required. Nevertheless, our proposal can also be extended to multi-level systems as shown in \ref{Transmon_Appen}. We like to emphasize that our system might have applications beyond quantum information processing. In particular, we have seen an improvement of excitation transfer between two nonidentical two-level systems when $g_p/\omega_{\rm{cav}}$ is increased, while one of them is incoherently pumped, thus providing the possibility of a minimal model for a photovoltaic cell. The extended study of this will be shown in our future article.\\

\section*{Acknowledgments}
We acknowledge fruitful discussion with Felipe Herrera. T.H.K. and L.-C.K. are supported by the National Research Foundation \& Ministry of Education, Singapore. S.A. acknowledges the financial support from Financiamiento Basal para Centros Cient\'ificos y Tecnol\'ogicos de Excelencia (Grant No. FB0807) and the Fondo Nacional de Desarrollo Cient\'ifico y Tecnol\'ogico (FONDECYT, Chile) under grant No. 1161018. G.R. acknowledges the support from the Fondo Nacional de Desarrollo Cient\'ifico y Tecnol\'ogico (FONDECYT, Chile) under grant No. 1150653.

\appendix

\section{Transmon-based implementation}
\label{Transmon_Appen}
Our proposal could be implemented in a realistic circuit QED setup by making use of a flux-qubit galvanically embedded inside a $\lambda/2$ microwave resonator for implementing the QRS. Two additional transmon circuits can be positioned at the resonator edges, where the resonator voltage is maximum. In this way, possible crosstalk between on-chip flux lines can be avoided. We also note that since transmons are many-level systems rather than qubits, so we need to carry out additional analysis in order to demonstrate our proposal. The Hamiltonian for a single transmon device reads
\begin{equation}
H_{T}=4E_C(N-N_g)^2-E_J\cos(\theta),
\label{HT}
\end{equation}    
where $E_C$ is the charging energy, $E_J$ is the Josephson energy, and $N_g$ is the effective offset charge of the device. Also, $N$ represents the number of Cooper pairs transferred to the superconducting island, and $\theta$ stands for the gauge-invariant phase difference across the Josephson junction.
\begin{figure}[b]
\centering
\includegraphics[scale=0.5]{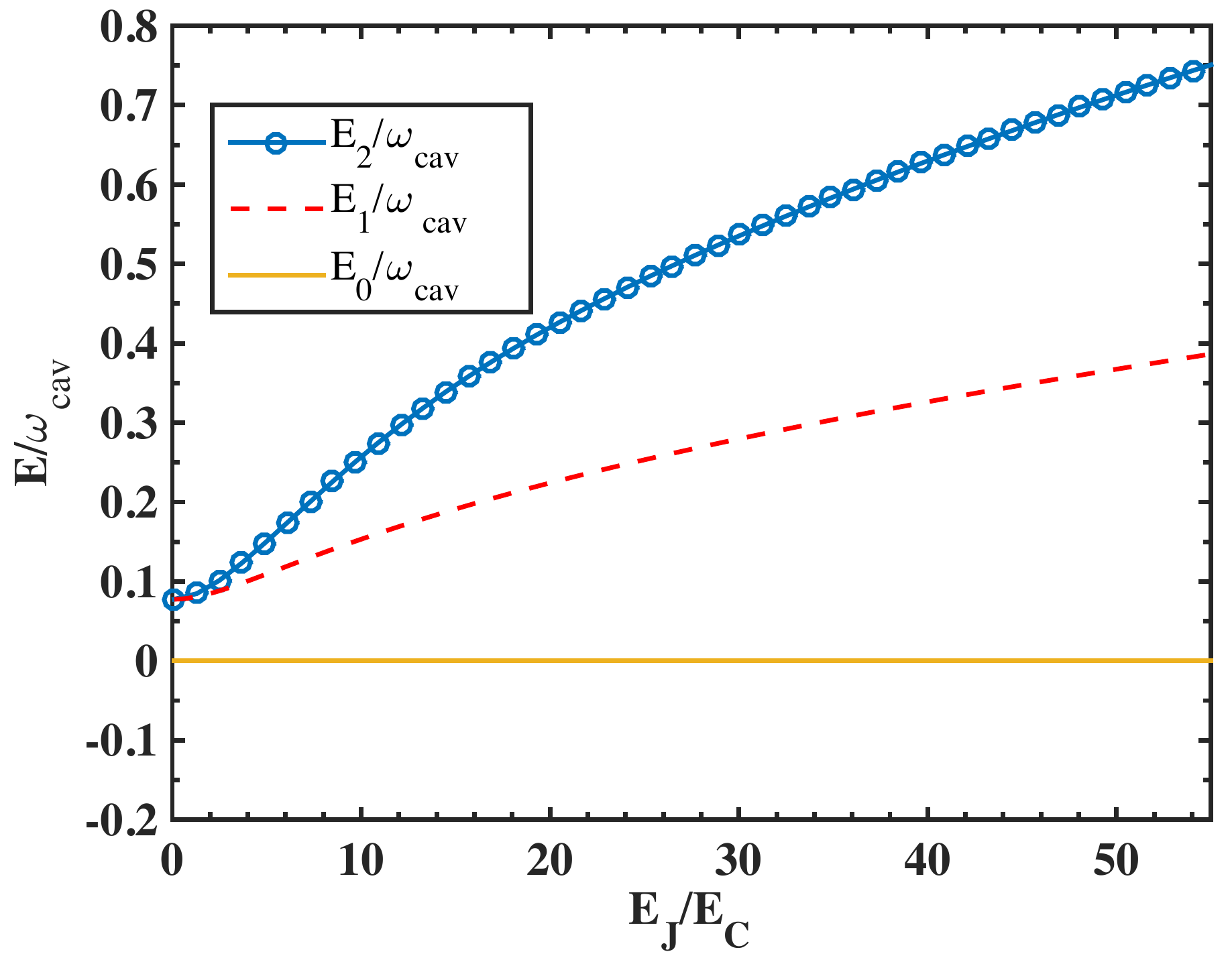}
\caption{Energy spectrum vs $E_J/E_C$ for the transmon device described by the Hamiltonian~(\ref{HT}). Here, the charging energy $E_C/\hbar=0.0194\omega_{\rm cav}$.} 
\label{FigSupp1}
\end{figure}
Figure~\ref{FigSupp1} shows the lowest three energy levels of the transmon as a function of the ratio $E_J/E_C$, and $N_g=0$. Here, we take realistic transmon charging energy $E_C/\hbar=2\pi\times0.31~{\rm GHz}$~\cite{PhysRevLett.112.070502}. Also, if we consider a cavity frequency $\omega_{\rm cav}=2\pi\times 16~{\rm GHz}$, which can be achieved with current circuit QED setups~\cite{Private}, we obtain $E_C/\hbar=0.0194\omega_{\rm cav}$. In this simulation we have fixed the ground state energy to the zero. Notice that for parameters of the QRS $\omega_p=\omega_{\rm cav}$ and $g_p=0.3\omega_{\rm cav}$, one can demonstrate that the frequency difference between the ground and excited state of the QRS is about $\omega_{10}=1.4\omega_{\rm cav}$. Also, at $E_J/E_C\approx 49$ ($E_J/\hbar=2\pi\times15.3~{\rm GHz}$~\cite{PhysRevLett.112.070502}) the absolute anharmonicity of the transmon is $\alpha=E_{10}-E_{21}=0.0223\omega_{\rm cav}=2\pi\times356.8~{\rm MHz}$.   
\begin{figure}[t]
\centering
\includegraphics[scale=0.5]{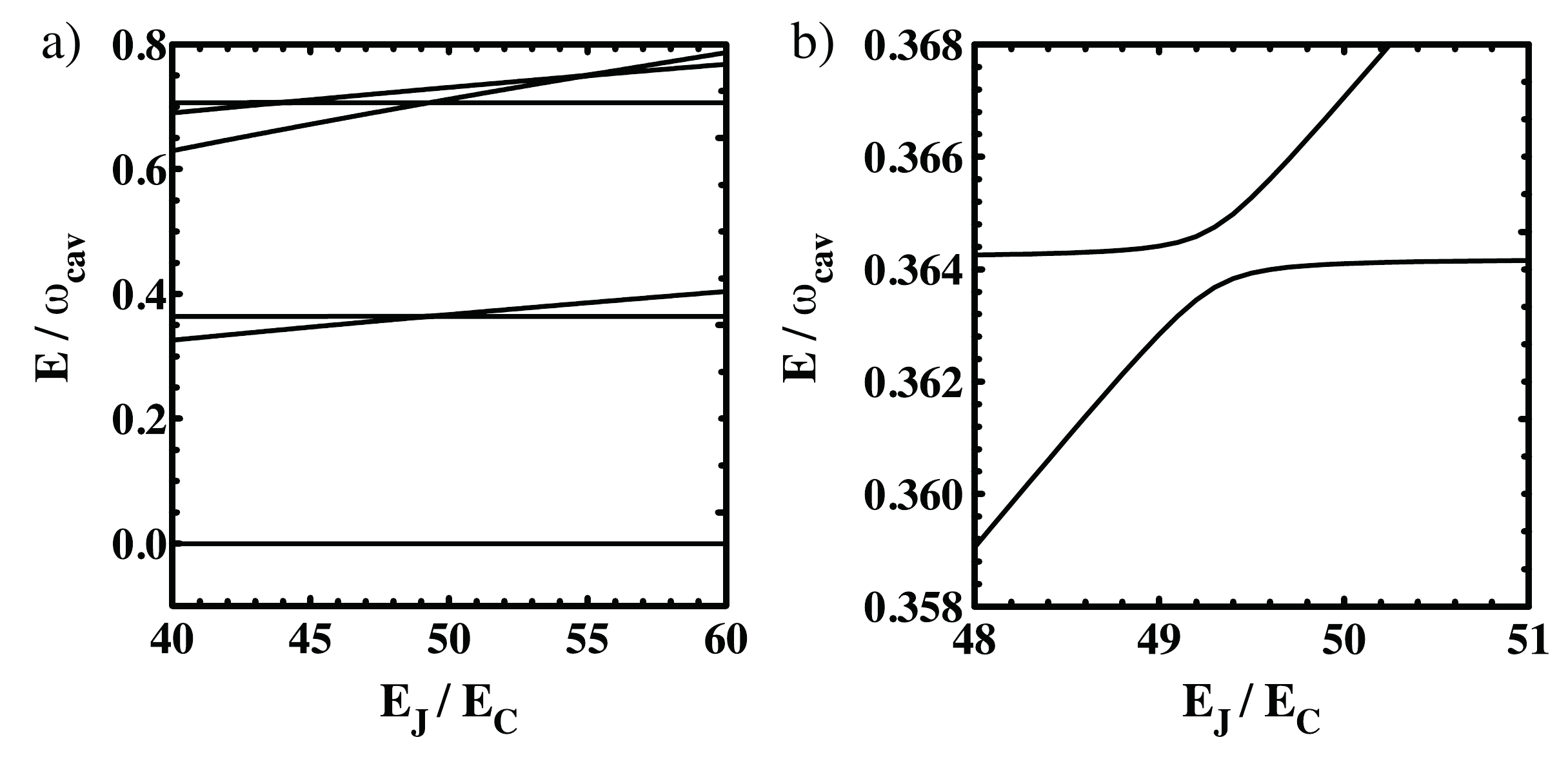}
\caption{a) Transmon- QRS-Trasmon energy spectrum for the Hamiltonian~(\ref{HTT}) vs the ratio $E_J/E_C$ for the transmon $2$. b) Enlarged first avoided level crossing in (a). Here, the parameters for transmon $1$ are fixed to $E_C=0.011\omega_{\rm cav}$, $E_J/E_C\approx 49$, and the parameters for the  QRS are $g_p/\omega_{\rm cav}=0.3$ and $\omega_p=\omega_{\rm cav}$.} 
\label{FigSupp2}
\end{figure} 

Based on the above results, we have performed simulations that include two nonidentical transmons coupled to the  QRS, and we truncate each transmon equation~(\ref{HT}) to its three lowest energy levels. The Hamiltonian that describes this situations reads
\begin{equation}
H= \sum^2_{\ell=1}\sum^2_{j=0}E^{(\ell)}_{j}\ket{j_\ell}\bra{j_\ell}+H_p+ \sum^{2}_{\ell=1}g_\ell N_\ell(b+b^{\dag}),
\label{HTT}
\end{equation}
where $E^{(\ell)}_{j}$ and $\ket{j_\ell}$ stand for the $j$th frequency and eigenstate associated with the $\ell$th transmon, respectively. $H_p$ is the Hamiltonian of the  QRS, $g_\ell$ and $N_\ell$ are the $\ell$th transmon- QRS coupling strength and Copper pairs number, respectively. Figure~\ref{FigSupp2} shows the results as a function of the ratio $E_J/E_C$ for transmon $2$, and for fixed ratio $E_J/E_C$ for the transmon $1$ taken from the simulation performed in figure~\ref{FigSupp1}. We see that at the resonance condition $E_J/E_C\approx 49$, the lowest avoided level crossing appears at $E/\omega_{\rm cav}\approx 0.3646$, see figure~\ref{FigSupp2}(b), which is below from the  QRS exicted state frequency, $\omega_{10}=1.4\omega_{\rm cav}$. Thus, the  QRS remains at its ground state. This resembles the avoided crossing that appears in figure~\ref{Fig4}(b) of the main text. Also, it shows evidence of the effective transmon-transmon interaction mediated by  QRS, such that the transmon-based implementation would allow us to simulate our parity-preserving light-matter system. This is analogous to the transmon-based implementation of effective qubit-qubit interaction in circuit QED setups~\cite{Majer:2007aa}. Notice that figure~\ref{FigSupp2}(a) also shows additional level crossings at higher energies that are caused by extra allowed transitions in the effective transmon-transmon interaction. \\

\section*{References}
\bibliographystyle{iopart-num}
\bibliography{Polariton_QST_IOPv2}
\end{document}